\documentclass[11pt]{amsart}
\usepackage{graphics}
\usepackage{hyperref}
\numberwithin{equation}{section}
\def \dd{{\rm d}}

\begin{document}

\title[]{Revisiting Weyl's calculation of the gravitational pull
in Bach's two-body solution}%
\author{Salvatore Antoci}%
\address{Dipartimento di Fisica ``A. Volta'' and INFM, Pavia, Italy}%
\email{Antoci@fisav.unipv.it}%
\author{Dierck-Ekkehard  Liebscher}%
\address{Astrophysikalisches Institut Potsdam, Potsdam, Germany}%
\email{deliebscher@aip.de}%
\author{Luigi Mihich}%
\address{Dipartimento di Fisica ``A. Volta'' and INFM, Pavia, Italy}%
\email{Mihich@fisav.unipv.it}%

%\thanks{}%
%subjclass{}%
%keywords{}%

%\date{}%
%\dedicatory{}%
%\commby{}%
% ----------------------------------------------------------------
\begin{abstract}
When the mass of one of the two
bodies tends to zero, Weyl's definition of the gravitational force
in an axially symmetric, static two-body solution can be given
an invariant formulation in terms of a force four-vector. The
norm of this force is calculated for Bach's two-body solution,
that is known to be in one-to-one correspondence with Schwarzschild's
original solution when one of the two masses $l$, $l'$ is made to
vanish. In the limit when, say,  $l'\rightarrow 0$, the norm of the force
divided by $l'$ and calculated at the position of the vanishing
mass is found to coincide with the norm of the acceleration of a
test body kept at rest in Schwarzschild's field. Both norms
happen thus to grow without limit when the test body (respectively
the vanishing mass $l'$) is kept at rest in a position closer and
closer to Schwarzschild's two-surface.
\end{abstract}
\maketitle
% ----------------------------------------------------------------
\section{Introduction}
It is well known since a long time (see {\it e.g.} (\cite{Rindler60}))
that a test body kept at rest in Schwarzschild's
gravitational field undergoes a four-acceleration whose norm
tends to infinity if the position of the body is closer and
closer to the Schwarzschild two-surface. However the existence of
this singular behaviour of the gravitational pull, despite the fact
that it can be given an invariant description, has not generally
aroused very much concern. By far greater attention has been
directed to the features of the world line followed by a test body
in free motion, and already in 1950 it has been shown
(\cite{Synge50}) that, once the singularity in the components of the
metric is removed through a coordinate transformation that
is aptly non regular at the Schwarzschild surface, a radial
timelike geodesic may reach the Schwarzschild singularity and
``cross it without a bump''. This early finding
by Synge has been the turning point for the program of analytic
extension \cite{Kruskal60},\cite{Szekeres60}.
Of course, geodesic motion has a fundamental role in general
relativity. Nevertheless, it is quite reassuring that the very notion
of the force exerted on a test body that is kept at rest, that has
played so fundamental a role in the development of physical
knowledge, does find a meaningful, {\it i.e.} invariant definition
in general relativity. It is less reassuring that the norm of that
force may be allowed to grow without limit at some surface in
the interior of a manifold meant to be a realistic model of physical
occurrences, without providing a justification for this
allowance in physical terms.\par The definition of the
gravitational force felt by a test body of unit mass kept
at rest in a static field  is connected to the geometric
definition of the four-acceleration by way of
hypothesis. It would be interesting to calculate this force in an
invariant way without availing of this hypothesis:
Einstein's equations alone should suffice for the task. In the
present paper the norm of the force exerted on a test body
in Schwarzschild's field is obtained by starting, in the footsteps of Weyl
\cite{Weyl19},\cite{BW22}, from a particular two-body solution
of Einstein's equations calculated in 1922 by R. Bach \cite{BW22}.

\section{Bach's solution for two ``point masses''}
While the spherically symmetric field of a ``Massenpunkt'' was
determined by K. Schwarzschild \cite{Schw16} soon after the
discovery of the field equations of general relativity
\cite{Einstein15b}, \cite{Hilbert15} \footnote{Schwarzschild
actually worked with the next-to-last version of the theory
\cite{Einstein15a}, whose covariance was limited to unimodular
transformations. As we shall see later, this fortuitous circumstance
had momentous consequences.} the class of axially symmetric, static
solutions, that could provide some indication about the
gravitational pull in a two-body system, was later found
\cite{Weyl17}, \cite{Levi-Civita19} by Weyl and by Levi-Civita.
Despite the nonlinear structure of the field equations, Weyl
succeeded in reducing the problem to quadratures through the
introduction of his ``canonical cylindrical coordinates''. Let
$x^0=t$ be the time coordinate, while $x^1=z$, $x^2=r$ are the
coordinates in a meridian half-plane, and $x^3=\varphi$ is the
azimuth of such a half-plane; then the line element of a static,
axially symmetric field {\it in vacuo} can be tentatively
written as:
\begin{equation}\label{2.1}
\dd s^2=e^{2\psi}dt^2-\dd\sigma^2,\;e^{2\psi}\dd\sigma^2
=r^2\dd\varphi^2+e^{2\gamma}(\dd r^2+\dd z^2);
\end{equation}
the two functions $\psi$ and $\gamma$ depend only on $z$ and $r$.
Remarkably enough, in the ``Bildraum'' introduced by Weyl $\psi$
fulfils the potential equation
\begin{equation}\label{2.2}
\Delta\psi=\frac{1}{r}\left\{\frac{\partial(r\psi_z)}
{\partial z}
+\frac{\partial(r\psi_r)}{\partial r}\right\}=0
\end{equation}
($\psi_z$, $\psi_r$ are the derivatives with respect to $z$ and to
$r$ respectively), while $\gamma$ is obtained by solving the system
\begin{equation}\label{2.3}
\gamma_z=2r\psi_z\psi_r,\;\gamma_r=r(\psi^2_r-\psi^2_z);
\end{equation}
due to the potential equation (\ref{2.2})
\begin{equation}\label{2.4}
\dd\gamma=2r\psi_z\psi_r\dd z+r(\psi^2_r-\psi^2_z)\dd r
\end{equation}
happens to be an exact differential.\par Schwarzschild's original
``Massenpunkt''solution \cite{Schw16} is recovered exactly when
$\psi$ is that solution of (\ref{2.2}) corresponding to the
Newtonian potential that one obtains if one segment of the
$z$-axis is covered by matter with constant mass
density \cite{Weyl17}. Let $2l$ be the coordinate length of this
segment, and let $r_1$, $r_2$ be the ``distances'', calculated in
the Euclidean way, of a point $P$ with canonical coordinates $z$, $r$
from the end points $P_1$ and $P_2$ of the segment, that lie on the
symmetry axis at $z=z_1$ and at $z=z_2=z_1-2l$ respectively.
One finds
\begin{equation}\label{2.5}
\psi=\frac{1}{2}\ln\frac{r_1+r_2-2l}{r_1+r_2+2l};\;
\gamma=\frac{1}{2}\ln\frac{(r_1+r_2)^2-4l^2}{4r_1r_2}.
\end{equation}
Through a coordinate transformation in the meridian half-plane one
proves the agreement with Schwarzschild's result, if $m$ is
substituted for $l$. We draw the attention of the reader on the
fact that the agreement occurs with Schwarzschild's original
solution \cite{Schw16}, not with what is called ``Schwarzschild
solution'' in all the textbooks, but is in fact the spherically
symmetric solution found by Hilbert through his peculiar choice
of the radial coordinate \cite{Hilbert17}. In fact by setting
\begin{equation}\label{2.6}
z-z_2=l(1-\cos{\vartheta}),
\end{equation}
one finds that the spatial part $\dd\sigma^2$ of the square of the
line element on the segment $P_2P_1$ becomes
\begin{equation}\label{2.7}
\dd\sigma^2=4l^2(\dd \vartheta^2+\sin^2{\vartheta}\dd\varphi^2),
\end{equation}
{\it i.e.} it coincides with the square of the line element of the
spherical two-surface that is present at $r=0$ in Schwarzschild's
true and original solution \cite{Schw16} of Einstein's equations. The
solution given by (\ref{2.5}) is in one-to-one correspondence with
that solution, and it does not contain  the ``interior region'' that
Hilbert could not help finding \cite{Hilbert17} due to the
particular way he kept in fixing the radial coordinate.\par
A static two-body solution is instead obtained if one assumes,
like Bach did \cite{BW22}, that the ``Newtonian potential''
$\psi$ is generated by matter that is present with constant
mass density on two segments of the symmetry axis, like the
segments $P_4P_3$ and $P_2P_1$ of Figure 1.
\begin{figure}[h]
\includegraphics{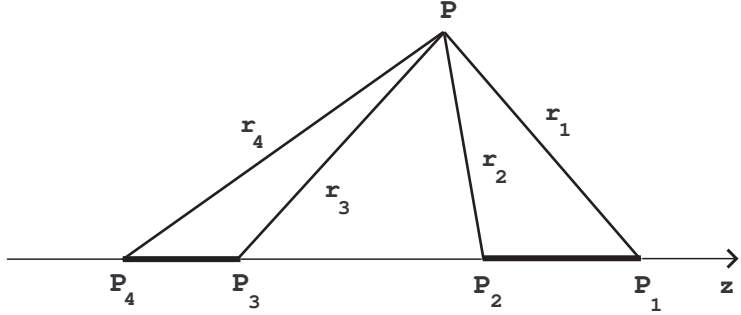}
\caption{Representation in the canonical $z$, $r$ half-plane of
the mass sources for Bach's two-body solution.
$r_4$, $r_3$ and $r_2$, $r_1$ are the ``distances'', calculated in
the Euclidean way, of a point $P$ from
the end points of the two segments endowed with mass.
$\overline{P_4P_3}=2l'$, $\overline{P_3P_2}=2d$, $\overline{P_2P_1}=2l$,
again in coordinate lengths.}
\end{figure}
We know already that the particular choice
\begin{equation}\label{2.8}
\psi=\frac{1}{2}\ln\frac{r_1+r_2-2l}{r_1+r_2+2l}
+\frac{1}{2}\ln\frac{r_3+r_4-2l'}{r_3+r_4+2l'},
\end{equation}
will produce a vacuum solution to Einstein's field equations
that reduces to Schwarzschild's original solution if one sets
either $l=0$ or $l'=0$. Of course, due to the nonlinearity of
(\ref{2.3}) one cannot expect that $\gamma$ will contain only the
sum of the contributions
\begin{equation}\label{2.9}
\gamma_{11}=\frac{1}{2}
\ln\frac{(r_1+r_2)^2-4l^2}{4r_1r_2},\;
\gamma_{22}=\frac{1}{2}
\ln\frac{(r_3+r_4)^2-4{l'}^2}{4r_3r_4},
\end{equation}
corresponding to the individual terms of the
potential (\ref{2.8}); a further term is present, that Bach
called $\gamma_{12}$, and reads
\begin{equation}\label{2.10}
\gamma_{12}
=\ln\frac{lr_4-(l'+d)r_1-(l+l'+d)r_2}{lr_3-dr_1-(l+d)r_2}+c,
\end{equation}
where $c$ is a constant.
Since $\gamma$ must vanish at the spatial infinity, it must be
$c=\ln[d/(l+l')]$. With this choice of the constant
one eventually finds \cite{BW22} that the line element of the two-body
solution is defined by the functions
\begin{eqnarray}\nonumber
e^{2\psi}=\frac{r_1+r_2-2l}{r_1+r_2+2l}
\cdot\frac{r_3+r_4-2l'}{r_3+r_4+2l'},\\\nonumber
e^{2\gamma}=\frac{(r_1+r_2)^2-4l^2}{4r_1r_2}
\cdot\frac{(r_3+r_4)^2-4{l'}^2}{4r_3r_4}\\\label{2.11}
\cdot\left(\frac{d(l'+d)r_1+d(l+l'+d)r_2-ldr_4}
{d(l'+d)r_1+(l+d)(l'+d)r_2-l(l'+d)r_3}\right)^2.
\end{eqnarray}
With these definitions for $\psi$ and $\gamma$ the line
element (\ref{2.1}) behaves properly at the spatial infinity
and is regular everywhere, except for the two segments $P_4P_3$,
$P_2P_1$ of the symmetry axis, where the sources of $\psi$
are located, and also for the segment $P_3P_2$, because there
$\gamma$ does not vanish as required, but takes the constant
value
\begin{equation}\label{2.12}
\Gamma=\ln{\frac{d(l+l'+d)}{(l+d)(l'+d)}},
\end{equation}
thus giving rise to the well known conical singularity.
\section{Weyl's analysis of the static two-body solutions}
Due to this lack of elementary flatness occurring on the segment
$P_3P_2$ the solution is not a true two-body solution;
nevertheless Weyl showed \cite{BW22} that a regular solution
could be obtained from it, provided that nonvanishing energy
tensor density $\mathbf{T}^k_i$ be allowed for in the space
between the two bodies. In this way an axial force $K$ is introduced,
with the evident function of keeping the two bodies at rest
\footnote{If a metric is static the definition of rest with respect
to that metric can be given in invariant form through the Killing
vectors.} despite their mutual gravitational attraction.
By providing a measure for $K$, Weyl provided a measure of the
gravitational pull. Let us recall here Weyl's analysis
\cite{Weyl19},\cite{BW22} of the axially symmetric, static
two-body problem.\par
In writing Einstein's field equations, we adopt henceforth Weyl's
convention for the energy tensor:
\begin{equation}\label{3.1}
R_{ik}-\frac{1}{2}g_{ik}R=-T_{ik}.
\end{equation}
Einstein's equations teach that, when the line element has the
expression (\ref{2.1}), $\mathbf{T}^k_i$ shall have the form
\begin{equation}\label{3.2}
\left(\begin{array}{llll}
\mathbf{T}^0_0&0&0&0\\
\\
0&\mathbf{T}^1_1&\mathbf{T}^1_2&0\\
\\
0&\mathbf{T}^2_1&\mathbf{T}^2_2&0\\
\\
0&0&0&\mathbf{T}^3_3
\end{array}\right)
\end{equation}
where
\begin{equation}\label{3.3}
\mathbf{T}^1_1+\mathbf{T}^2_2=0.
\end{equation}
By introducing the notation
\begin{equation}\label{3.4}
\mathbf{T}^3_3=r\varrho',\;\mathbf{T}^0_0=r(\varrho+\varrho'),
\end{equation}
Einstein's equations can be written as:
\begin{equation}\label{3.5}
\Delta\psi=\frac{1}{2}\varrho,\;
\frac{\partial^2\gamma}{\partial z^2}
+\frac{\partial^2\gamma}{\partial r^2}
+\left\{\left(\frac{\partial\psi}{\partial z}\right)^2
+\left(\frac{\partial\psi}{\partial r}\right)^2\right\}
=-\varrho';
\end{equation}
\begin{equation}\label{3.6}
\mathbf{T}^1_1=-\mathbf{T}^2_2
=\gamma_r-r(\psi^2_r-\psi^2_z),\;
-\mathbf{T}^2_1=-\mathbf{T}^1_2=\gamma_z-2r\psi_r\psi_z.
\end{equation}
Weyl shows that $\varrho$ must be interpreted as mass density in
the canonical space. To this end he considers the mass density
distribution sketched in Figure 2, where $\varrho$ is assumed to be
nonvanishing only in the shaded regions labeled $1$ and $2$.
\begin{figure}[h]
\includegraphics{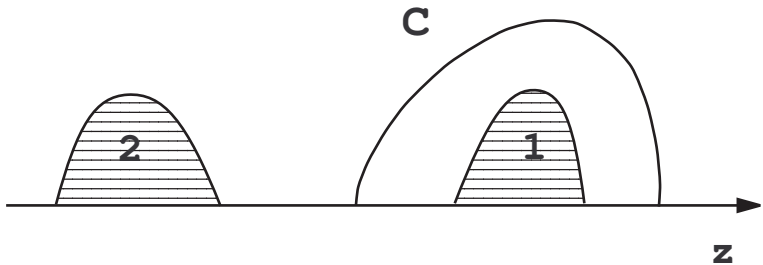}
\caption{Representation in the canonical $z$, $r$ half-plane
of extended mass sources of a two-body solution.}
\end{figure}
According to (\ref{3.5}) the potential $\psi$ corresponding to this
mass distribution can be uniquely split in two terms $\psi_1$ and
$\psi_2$, such that $\psi_1$ is a potential function that vanishes
at infinity and is everywhere regular outside the region $1$,
while $\psi_2$ behaves in the same way outside the region $2$. The
asymptotic forms of $\psi_1$ and $\psi_2$ are such that
\begin{equation}\label{3.7}
e^{2\psi_1}=1-\frac{m_1}{R}+...,\;
e^{2\psi_2}=1-\frac{m_2}{R}+...
\end{equation}
where the mass coefficients $m_1$ and $m_2$ are given by the
integral $\int{\varrho \dd V}=2\pi\int{\varrho r\dd r\dd z}$, performed
in the canonical space and extended to the appropriate shaded
region. Outside the shaded regions one has $\varrho=0$, but
there shall be some region between the bodies, let us call it $L'$,
where $\varrho=0$ but $\mathbf{T}^k_i\neq 0$, since in a static
solution of general relativity the gravitational pull shall be
counteracted in some way. Weyl's procedure for determining
$\mathbf{T}^k_i$ in $L'$ is the following. Suppose that $\mathbf{T}^k_i=0$
outside a simply connected region $L$ that includes both material bodies.
Since $\psi$ is known there, we can avail of (\ref{2.4}), together with
the injunction that $\gamma$ vanish at infinity, to determine $\gamma$
uniquely outside $L$. Within $L'$ we can choose $\gamma$
arbitrarily, provided that we ensure the regular connection with
the vacuum region and the regular behaviour on the axis,
{\it i.e.} $\gamma$ vanishing there like $r^2$. Since $\psi$ is
known in $L'$ and $\gamma$ has been chosen as just shown, we can
use equations (\ref{3.5}) and (\ref{3.6}) to determine
$\mathbf{T}^k_i$ there.\par
If the material bodies include each one a segment of the axis, just as it
occurs in Fig. 2, the force $K$ directed along the $z$ axis,
with which the stresses in $L'$ contrast the gravitational pull
can be written as
\begin{equation}\label{3.8}
K=2\pi\int_C(\mathbf{T}^2_1 \dd z-\mathbf{T}^1_1 \dd r);
\end{equation}
the integration path is along a curve $C$, like the one drawn in
Fig. 2, that separates the two bodies in the meridian half-plane;
the value of the integral does not depend on the precise position
of $C$ because, as one gathers from the definitions (\ref{3.5}),
(\ref{3.6}):
\begin{equation}\label{3.9}
\mathbf{T}^1_{1,1}+\mathbf{T}^2_{1,2}=0
\end{equation}
in the region $L'$. Since the region of the meridian half-plane
where $\varrho=0$ is simply connected, by starting from $\psi$
and from the vacuum equation (\ref{2.4}), now rewritten as:
\begin{equation}\label{3.10}
\dd\gamma^*=2r\psi_z\psi_r\dd z+r(\psi^2_r-\psi^2_z)\dd r
\end{equation}
one can uniquely define there the function $\gamma^*$ that
vanishes at the spatial infinity. In all the
parts of the $z$ axis where $\varrho=0$ it must be $\gamma^*_z=0$,
$\gamma^*_r=0$, hence $\gamma^*=const.$, $\gamma^*_r=0$. In
particular, in the parts of the axis that go to infinity one shall
have $\gamma^*=0$; let us call $\Gamma^*$ the constant value
assumed instead by $\gamma^*$ on the segment of the axis lying
between the two bodies. The definitions (\ref{3.6}) can now be
rewritten as:
\begin{equation}\label{3.11}
\mathbf{T}^1_1=-\mathbf{T}^2_2=\gamma_r-\gamma^*_r,\;
-\mathbf{T}^2_1=-\mathbf{T}^1_2=\gamma_z-\gamma^*_z,
\end{equation}
and the integral of (\ref{3.8}) becomes
\begin{equation}\label{3.12}
\int_C(\mathbf{T}^2_1 \dd z-\mathbf{T}^1_1 \dd r)
=\int_C{(\gamma^*_z-\gamma_z)\dd z+(\gamma^*_r-\gamma_r)\dd r}
=\int_C \dd(\gamma^*-\gamma).
\end{equation}
Since $\gamma$ vanishes on the parts of the $z$ axis where
$\varrho=0$, the force $K$ that holds the bodies at rest
despite the gravitational pull shall be
\begin{equation}\label{3.13}
K=-2\pi\Gamma^*
\end{equation}
with Weyl's definition (\ref{3.1}) of the energy tensor. When the
mass density $\varrho$ has in the canonical space the particular
distribution considered by Bach and drawn in Fig. 1,
$\Gamma^*$ is equal to $\Gamma$ as defined by (\ref{2.12}).
The measure of the gravitational pull with which the two ``material
bodies'' of this particular solution attract each other therefore turns
out to be
\begin{equation}\label{3.14}
K=2\pi\ln{\frac{(d+l)(d+l')}{d(d+l+l')}}
\end{equation}
in Weyl's units. This expression agrees with the Newtonian value
when $l$ and $l'$ are small when compared to $d$, as expected.

\section{From Weyl's $K$ to a ``quasi'' force four-vector $k_i$}
Despite its mathematical beauty, Weyl's definition of the
gravitational pull for an axially symmetric, static two-body
solution appears associated without remedy to the adoption
of the canonical coordinate system. It is however possible to
obtain through Weyl's definition of $K$, given by (\ref{3.8}),
a ``quasi'' four-vector $k_i$. In fact that expression can be
rewritten as
\begin{equation}\label{4.1}
K=\int_{\Sigma}\mathbf{T}^l_1 \dd f^*_{0l}
\equiv\frac{1}{2}\int_{\Sigma} T^l_1\epsilon_{0lmn}\dd f^{mn},
\end{equation}
where $\epsilon_{klmn}$ is Levi-Civita's totally antisymmetric
tensor and $\dd f^{mn}$ is the element of the two-surface $\Sigma$
generated by the curve $C$ through rotation around the symmetry
axis. Since the metric is static it is possible to define invariantly a
timelike Killing vector $\xi^k_{(t)}$ that correspond, in Weyl's
canonical coordinates, to a unit coordinate time translation. Therefore
(\ref{4.1}) can be rewritten as
\begin{equation}\label{4.2}
K=\frac{1}{2}\int_{\Sigma}\xi^k_{(t)}T^l_1\epsilon_{klmn}\dd f^{mn}
\end{equation}
by still using the canonical coordinates. Now the integrand is
written as the ``$1$'' component of the infinitesimal
covariant four-vector
\begin{equation}\label{4.3}
\xi^k_{(t)}T^l_i\epsilon_{klmn}\dd f^{mn},
\end{equation}
but of course in general the expression
\begin{equation}\label{4.4}
k_i=\frac{1}{2}\int_{\Sigma}
\xi^k_{(t)}T^l_i\epsilon_{klmn}\dd f^{mn}
\end{equation}
will not be a four-vector, because the integration over $\Sigma$ spoils
the covariance. When evaluated in canonical coordinates, the
nonvanishing components of $k_i$ are $k_1=K$ and
\begin{equation}\label{4.5}
k_2=2\pi\int_C(\mathbf{T}^2_2 \dd z-\mathbf{T}^2_1 \dd r)
=2\pi\int_C{(\gamma^*_r-\gamma_r)\dd z-(\gamma^*_z-\gamma_z)\dd r},
\end{equation}
that however must vanish too, if $k_i$ has to become a four-vector
defined on the symmetry axis. But, as one sees from
Weyl's analysis, we are at freedom to choose $\mathbf{T}^k_i$ in
$L'$ as nonvanishing only in a tube with a very small~
\footnote{This kind of procedure has been
used to derive the equations of motion even for structured particles
by Einstein, Infeld and Hoffman \cite{EIH38} and by Fock and
Papapetrou (see \cite{Papapetrou51}).} yet finite coordinate
radius that encloses in its interior the segment of the symmetry
axis lying between the bodies; moreover, we can freely set
$\gamma_z=\gamma^*_z$ within the tube. Under these conditions
the second term of the integral (\ref{4.5}) just vanishes, while
the first one shall be very small, since the regularity of the surface
$\Sigma$ requires that the curve $C$ approach the symmetry axis at a
straight angle in canonical coordinates. By properly choosing
$\mathbf{T}^k_i$ we thus succeed in providing through equation
(\ref{4.4}) a quasi four-vector $k_i$ whose components, written
in Weyl's canonical coordinates, reduce in approximation to $(K,0,0,0)$.
\section{The norm of the force in Bach's solution when $2l'\rightarrow 0$}
Having defined, with the above caveats, the quasi four-vector $k_i$
along the segment of the symmetry axis between the two bodies,
we can use its ``quasi'' norm to provide a measure of the force
that opposes the gravitational pull. In the case of Bach's two-body
solution, whose line element is defined in canonical coordinates
by (\ref{2.1}) and (\ref{2.11}), that quasi norm reads
\begin{equation}\label{5.1}
k\equiv(-k^ik_i)^{1/2}
=2\pi\ln\frac{(d+l)(d+l')}{d(d+l+l')}
\cdot\left[\frac{r_1-2l}{r_1}
\cdot\frac{r_4-2l'}{r_4}\right]^{1/2}
\end{equation}
when measured in Weyl's units at a point of the symmetry
axis for which $z_3<z<z_2$. At variance with the behaviour of $K$,
the quasi norm $k$ depends on $z$, due to the term of (\ref{5.1})
enclosed within the square brackets, that comes from $e^{2\psi}$.
Let us evaluate this quasi norm divided by $l'$ when $l'\rightarrow 0$,
namely, the coefficient of the linear term in the McLaurin
series expansion of $k$ with respect to $l'$. Since $\Gamma^*$,
now defined by the right hand side of (\ref{2.12}),
tends to zero when $l'\rightarrow 0$, while performing this limit one can
also send to zero the radius of the very narrow tube considered in the
previous section. Therefore $k_i$ can become a true four-vector and $k$
can become a true norm in the above mentioned limit. With this
proviso one finds the invariant result
\begin{equation}\label{5.2}
\lim_{l'\rightarrow 0}\left[\frac{k}{l'}\right]
=\left[\frac{\partial k}{\partial l'}\right]_{l'=0}
=\frac{2\pi l}{d(d+l)}\left(\frac{r_1-2l}{r_1}\right)^{1/2}.
\end{equation}
When $l'\rightarrow 0$ the line element of Bach's solution
with two bodies tends to the line element defined by (\ref{2.1}) and
(\ref{2.5}), that is in one-to-one correspondence with the
line element of Schwarzschild's original solution \cite{Schw16}.
Therefore the scalar quantity $[\partial k/\partial l']_{l'=0}$
evaluated at $P_3$ shall be the norm of the force per unit mass
exerted by Schwarzschild's gravitational field on a test
particle kept at rest at $P_3$. Its value is obtained by substituting
$2d+2l$ for $r_1$ in (\ref{5.2}). One finds
\begin{equation}\label{5.3}
\left(\lim_{l'\rightarrow 0}\left[\frac{k}{l'}\right]\right)_{z=z_3}
=\frac{8\pi l}{(2d+2l)^{3/2}(2d)^{1/2}}.
\end{equation}
If one solves Schwarzschild's problem in spherical polar coordinates $r$,
$\vartheta$, $\varphi$, $t$ with three unknown functions of $r$,
{\it i.e.} without fixing the radial coordinate, like Combridge and Janne
did long ago \cite{Combridge23},\cite{Janne23}, one ends up to write
de~Sitter's line element \cite{de Sitter16}
\begin{equation}\label{5.4}
\dd s^2=-\exp{\lambda}\dd r^2
-\exp{\mu}[r^2(\dd\vartheta^2+\sin^2{\vartheta}\dd\varphi^2)]
+\exp{\nu}\dd t^2
\end{equation}
in terms of one unknown function $f(r)$. In fact $\lambda$,
$\mu$, $\nu$ are defined through this arbitrary function $f(r)$
and through its derivative ${f'}(r)$ as follows:
\begin{eqnarray}\label{5.5}
\exp{\lambda}=\frac{{f'}^2}{1-2m/f},\\\label{5.6}
\exp{\mu}=\frac{f^2}{r^2},\\\label{5.7}
\exp{\nu}=1-2m/f.
\end{eqnarray}
Here $m$ is the mass constant; of course the arbitrary function
$f$ must have the appropriate behaviour as $r\rightarrow\infty$.
Schwarzschild's original solution \cite{Schw16} is eventually recovered
\cite{Abrams79},\cite{Abrams89} by requiring that $f$ be a monotonic
function of $r$ and that $f(0)=2m$. Let us imagine that a test body be
kept at rest in this field. With our symmetry-adapted coordinates, its
world line shall be invariantly specified by requiring that the spatial
coordinates $r$, $\vartheta$, $\varphi$ of the test body be constant
in time. If
\begin{equation}\label{5.8}
\alpha=(-a_ia^i)^{1/2}
\end{equation}
is the norm of the acceleration four-vector
\begin{equation}\label{5.9}
a^i\equiv\frac{\dd u^i}{\dd s}+\Gamma^i_{kl}u^ku^l
\end{equation}
along the world line of the test body, one finds:
\begin{equation}\label{5.10}
\alpha=\frac {m}{f^{3/2}(f-2m)^{1/2}}.
\end{equation}
This norm is assumed by way of hypothesis to be equal to
the norm of the force per unit mass needed for constraining the
test particle to follow a world line of rest despite the
gravitational pull of the Schwarzschild field \cite{Rindler60}.
The consistency of the hypothesis with Einstein's theory requires
that $\alpha$ be equal to the scalar quantity
$[\partial k/\partial l']_{l'=0,\;z=z_3}$
that provides the norm of the force per unit mass for Bach's
solution in the test particle limit $l'\rightarrow 0$.
This is indeed the case, since the functional dependence of
(\ref{5.3}) on the mass parameter $l$ and on the coordinate distance
$2d+2l$ is the same as the functional dependence of (\ref{5.10})
on the mass parameter $m$ and on the function $f(r)$ with
$f(0)=2m$ introduced above. The extra constant $8\pi$ appearing in
(\ref{5.3}) is just due to Weyl's adoption of the definition (\ref{3.1})
of the energy tensor.\par For Schwarzschild's field, the
definition of the norm of the force exerted on a test particle at rest
obtained through the acceleration four-vector and the independent definition
through the force that, in Bach's two-body solution, $\mathbf{T}^k_i$
must exert to keep the masses at rest when $l'\rightarrow 0$ lead to
one and the same result. In particular, both definitions show
that the norm of the force per unit mass grows without limit as the
test particle is kept at rest in a position closer and closer to
Schwarzschild's two-surface.
\break

\newpage

\begin{thebibliography}{}

\bibitem{Rindler60} Rindler, W., Phys. Rev. {\bf 119} (1960)
2082.

\bibitem{Synge50} Synge, J.L., Proc. R. Irish Acad. {\bf 53A}
(1950) 83.

\bibitem{Kruskal60} Kruskal, M.D., Phys. Rev. {\bf 119} (1960)
1743.

\bibitem{Szekeres60} Szekeres, G., Publ. Math. Debrecen
{\bf 7} (1960) 285.

\bibitem{Weyl19} Weyl, H., Ann. d. Phys. {\bf59} (1919) 185.

\bibitem{BW22} Bach, R. and Weyl, H., Math. Zeitschrift {\bf 13}
(1922) 134.

\bibitem{Schw16} Schwarzschild, K., Sitzungsber. Preuss.
Akad. Wiss., Phys. Math. Kl. 1916, 189 (communicated 13 Jan.
1916).

\bibitem{Einstein15b} Einstein, A., Sitzungsber. Preuss.
Akad. Wiss., Phys. Math. Kl. 1915, 844 (communicated 25 Nov.
1915).

\bibitem{Hilbert15} Hilbert, D., Nachr. Ges. Wiss. G\"ottingen,
Math. Phys. Kl. 1915, 395 (communicated 20 Nov. 1915).

\bibitem{Einstein15a} Einstein, A., Sitzungsber. Preuss.
Akad. Wiss., Phys. Math. Kl. 1915, 778 (communicated 11 Nov. 1915).

\bibitem{Weyl17} Weyl, H., Ann. Phys. (Leipzig) {\bf 54}
(1917) 117.

\bibitem{Levi-Civita19} Levi-Civita, T., Rend. Acc. dei Lincei,
{\bf 28} (1919) 3.

\bibitem{Hilbert17} Hilbert, D., Nachr. Ges. Wiss. G\"ottingen,
Math. Phys. Kl. 1917, 53.

\bibitem{EIH38} Einstein, A., Infeld, L. and Hoffmann, B., Ann.
Math. {\bf 39} (1938) 65.

\bibitem{Papapetrou51} Papapetrou, A., Proc. Phys. Soc. {\bf A64}
(1951) 57.

\bibitem{Combridge23} Combridge, J.T., Phil. Mag. {\bf 45}
(1923) 726.

\bibitem{Janne23} Janne, H., Bull. Acad. R. Belg. {\bf 9}
(1923) 484.

\bibitem{de Sitter16} de Sitter, W., Month. Not. R. Astr.
Soc. {\bf 76} (1916) 699.

\bibitem{Abrams79} Abrams, L.S., Phys. Rev. D {\bf 20} (1979)
2474.

\bibitem{Abrams89} Abrams, L.S., Can. J. Phys. {\bf 67} (1989)
919.

\end{thebibliography}
\end{document}